\documentclass[10pt,letterpaper]{article}
\usepackage[top=0.85in,left=2.75in,footskip=0.75in]{geometry}

\usepackage{amsmath,amssymb}

\usepackage{changepage}

\usepackage{textcomp,marvosym}

\usepackage{cite}

\usepackage{nameref,hyperref}

\usepackage[right]{lineno}

\usepackage[nopatch=eqnum]{microtype}
\DisableLigatures[f]{encoding = *, family = * }

\usepackage{array}

\newcolumntype{+}{!{\vrule width 2pt}}

\newlength\savedwidth

\newcommand\thickhline{\noalign{\global\savedwidth\arrayrulewidth\global\arrayrulewidth 2pt}%
\hline
\noalign{\global\arrayrulewidth\savedwidth}}

\raggedright
\usepackage[aboveskip=1pt,labelfont=bf,labelsep=period,justification=raggedright,singlelinecheck=off]{caption}

\makeatletter
\renewcommand{\@biblabel}[1]{\quad#1.}
\makeatother

\usepackage{lastpage,fancyhdr,graphicx}
\usepackage{epstopdf}
\fancyheadoffset[L]{2.25in}
\fancyfootoffset[L]{2.25in}
\usepackage{algpseudocode,algorithm,lipsum,xspace}

\newcommand{\ie}{\emph{i.e.},\xspace}
\newcommand{\eg}{\emph{e.g.},\xspace}

\newcommand{\cmp}[1]{\mathcal{O}(#1)}
\newcommand{\avg}[1]{\langle#1\rangle}
\newcommand{\pval}{p\mbox{-value}}

\renewcommand{\eqref}[1]{Eq~(\ref{eq:#1})\xspace}
\newcommand{\figref}[1]{Fig~\ref{fig:#1}\xspace}
\newcommand{\tblref}[1]{Table~\ref{tbl:#1}\xspace}
\renewcommand{\algref}[1]{Algorithm~\ref{alg:#1}\xspace}

\begin{document}
\vspace*{0.2in}

\begin{flushleft}
{\Large
\textbf\newline{Computing well-balanced spanning trees of unweighted networks} 
}
\newline
\\
Lovro Šubelj\textsuperscript{1,2*}
\\
\bigskip
\textbf{1} University of Ljubljana, Faculty of Computer and Information Science, Ljubljana, Slovenia
\\
\textbf{2} University of Ljubljana, Faculty of Social Sciences, Ljubljana, Slovenia
\\
\bigskip

* lovro.subelj@fri.uni-lj.si

\end{flushleft}
\section*{Abstract}
A spanning tree of a network or graph is a subgraph that connects all nodes with the least number or weight of edges. The spanning tree is one of the most straightforward techniques for network simplification and sampling, and for discovering its backbone or skeleton. Prim's algorithm and Kruskal's algorithm are well-known algorithms for computing a spanning tree of a weighted network, and are therefore also the default procedure for unweighted networks in the most popular network libraries. In this paper, we empirically study the performance of these algorithms on unweighted networks and compare them with different priority-first search algorithms. We show that the structure of a network, such as the distances between the nodes, is better preserved by a simpler algorithm based on breadth-first search. The spanning trees are also most compact and well-balanced as measured by classical graph indices. We support our findings with experiments on synthetic graphs and more than a thousand real networks, and demonstrate practical applications of the computed spanning trees. We conclude that if a spanning tree is to maintain the structure of an unweighted network, the breadth-first search algorithm should be the preferred choice, and it should be implemented as such in network libraries.


\section*{Introduction}
\label{sec:intro}

Networks or graphs have become a popular tool for analyzing complex real-world systems~\cite{New18c}. Examples include predicting the spread of contagious viruses~\cite{TBPRBGPCV12}, the study of the interactome of species~\cite{ZSFL19}, understanding the structure of science~\cite{FBBEHMPRSUVWWB18} and outreach of online social connections~\cite{BBRUV12}. The size of today's networks is often in millions of nodes and edges, with the largest networks being the WWW with more than a trillion web pages and the human brain with nearly a hundred billion neurons. As a result, the size of real networks makes many practical applications computationally very challenging.

Techniques to alleviate this issue include network simplification or sampling~\cite{LF06,HLMSW16,BSB17} and revealing the so-called network backbone or skeleton~\cite{GTB12,CN17,Sub18a,SCR21}. These approaches try to reduce the size of a network in a way that the network still retains many of its structural properties. One of the most straightforward ways to simplify a network is to compute its spanning tree~\cite{Bol98,New18c}, which is a subgraph connecting all the nodes of a network with the minimum number of edges. Spanning trees retain the connectivity of networks and possibly other structural properties, and have gained considerable interest in recent years~\cite{Pop20,DBB22,HM24,Dho24,RVJ24}. In the case of weighted networks, one usually aims to compute the minimum spanning tree, which is a subgraph connecting all the nodes with the minimum total weight of the edges. In the case of unweighted networks, any spanning tree is, in fact, a ``minimum'' spanning tree.

Prim's and Kruskal's algorithms are well-known algorithms for computing a minimum spanning tree of a weighted network~\cite{Bol98,New18c}. Although the algorithms were primarily developed for weighted networks, they can be readily applied to unweighted networks where each edge has the same weight. This is also the default procedure in the most popular network analysis libraries {\it NetworkX}, {\it igraph}, and {\it graph-tool}~\cite{AA25}. However, the performance of these algorithms has not been sufficiently studied for unweighted networks, which are much more common in practical applications due to their simplicity~\cite{Abd11}. In particular, there exist no theoretical guarantees for the algorithms, and neither does the literature provide any empirical comparison on large-scale networks.

A fundamental result of metric embedding theory, Bourgain’s theorem~\cite{Bou85}, states that any finite metric space can be embedded into a tree metric with $\cmp{\log n}$ distortion. This suggests that while spanning trees provide a reasonable approximation of network structure, they inevitably introduce some bias that must be evaluated empirically.

\begin{figure}[t]
\centering\vskip8pt\includegraphics[width=\textwidth]{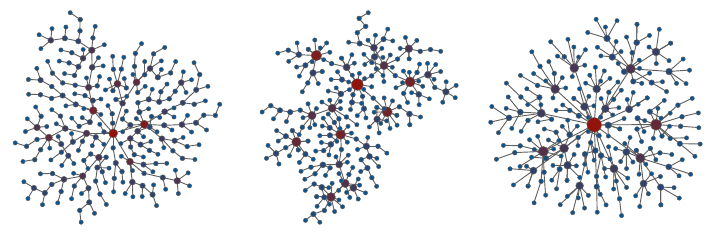}\vskip8pt%
\caption{{\bf Wiring diagrams of spanning trees of a small random graph.}
The spanning trees were computed with Prim's algorithm ({left}), Kruskal's algorithm ({middle}) and the breadth-first search algorithm ({right}). The size of the nodes is proportional to their degree, while the layouts were computed with the Large Graph Layout algorithm~\cite{ADWM04}.}
\label{fig:trees}
\end{figure}

In this paper, we apply the algorithms to different synthetic graphs and more than a thousand real networks, and compare them to different priority-first search algorithms. We show that the structure of unweighted networks is best preserved by an algorithm using the breadth-first search node traversal. More precisely, spanning trees computed with the breadth-first search algorithm best preserve the distances between the nodes of a network. Recall that other standard network measures, such as the average node degree or clustering coefficient~\cite{WS98}, are fixed by construction. The spanning trees are also most compact and well-balanced  as measured by Wiener's index~\cite{Wie47} and Sackin's index~\cite{Sac72}. 
Such structural properties are often desirable in practical applications such as network visualization (\figref{trees}). Furthermore, the breadth-first search algorithm can be up to $20$ times faster and uses less memory than alternatives.

The rest of the paper is structured as follows. In the following section, we first describe different algorithms for computing a spanning tree of a network, as well as different indices of tree balance and compactness. Next, we analyze the structure of spanning trees of synthetic graphs and then the structure of spanning trees of real networks. Finally, we demonstrate practical applications of spanning trees of networks. 

\section*{Methods}
\label{sec:methods}

Let a network be represented by an undirected connected graph $G=(V,E)$, where $V$ denotes the set of nodes of $G$ and $E$ denotes the set of edges of $G$. Where necessary to make explicit that these sets represent the graph $G$, we write $V_G$ and $E_G$. The number of nodes equals $n=|V|$ and the number of edges equals $m=|E|$. We denote the average node degree as $\avg{k}=2m/n$. Furthermore, let $d_{ij}$ be the distance between the nodes $i,j\in V$, defined as the number of edges in a shortest path between the nodes $i$ and $j$. Since the graph is undirected and connected, $d_{ij}=d_{ji}$ and $d_{ij}<\infty$. Therefore, the average distance between the nodes equals $\avg{d}=\frac{2}{n(n-1)}\sum_{i<j}d_{ij}$ and the maximum distance or diameter equals $d_{\rm max}=\max_{i<j}d_{ij}$. In order to measure the variability of the distances between the nodes, we also define the coefficient of variation as $c_d=\sigma_d/\avg{d}$, where $\sigma_d$ is the standard deviation of distances. The coefficient of variation $c_d$ is a measure of the dispersion of a probability distribution, where the distributions with $c_d<1$ are considered low-variance distributions, while those with $c_d>1$ are considered high-variance distributions.

Below, we describe different algorithms for computing a spanning tree of an undirected connected graph. In the case of a disconnected graph consisting of more than one connected component, the algorithms should be applied to each of the connected components separately. Implementation of these algorithms is provided by most standard network analysis libraries. For a more extensive discussion on the differences between the algorithms, we refer the reader to classical graph theory literature~\cite{Bol98} or network science literature~\cite{New18c,Bar16}.

\subsection*{Prim's algorithm}

Prim's algorithm for computing a spanning tree $T$ of an undirected connected graph $G$ operates as follows (see~\algref{prim}). First, the algorithm selects a random seed node $i\in V_G$ from graph $G$ and adds it to an empty tree $T$ (lines 2, 3). The node $i$ serves as a starting point for computing the spanning tree $T$. Then, on each step of the algorithm (lines 4-8), a random edge $\{i,j\}\in E_G$ from graph $G$ is selected that leads from a node $i\in V_T$ already in the tree $T$ to a node $j\notin V_T$ not yet in the tree $T$ (line 5). Both node $j$ and edge $\{i,j\}$ are added to the tree $T$ (lines 6, 7). Finally, when there is no further node $i\in V_G$ in graph $G$ such that node $i\notin V_T$ is not already in the tree $T$, the algorithm stops (line 4). At this point, the tree $T$ is a spanning tree of graph $G$ (line 9).

\begin{algorithm}[h]
	{\small\caption{\label{alg:prim}Prim's algorithm}
	\begin{algorithmic}[1]
		\Require undirected graph $G$
		\Ensure spanning tree $T$
		\State $T\gets$ empty graph
		\State $i\gets\Call{Random}{i\in V_G}$ \Comment{{\small Random seed node.}}
		\State add node $i$ to $V_T$ \Comment{{\small Add selected seed node.}}
		\While{$\exists i\in V_G: i\notin V_T$} \Comment{{\small There exists non-visited node?}}
			\State $\{i,j\}\gets\Call{Random}{\{i,j\}\in E_G: i\in V_T\wedge j\notin V_T}$ \Comment{{\small Edge to non-visited node.}}
			\State add node $j$ to $V_T$ \Comment{{\small Add non-visited node.}}
			\State add edge $\{i,j\}$ to $E_T$ \Comment{{\small Add selected edge.}}
		\EndWhile
		\State \Return $T$
	\end{algorithmic}}
\end{algorithm}

Prim's algorithm is nondeterministic and can compute different spanning trees. The actual spanning tree depends on a random selection of the seed node (line 2) and on a random selection of the edges to expand the tree (line 5). The latter is most efficiently implemented by rejection sampling over an array list of edges to non-visited nodes. For simplicity, we do not make these computations explicit in~\algref{prim}.

Assume that the graph is represented with an adjacency list. In the case of weighted graphs, the time complexity of Prim's algorithm implemented with a Fibonacci heap is $\cmp{m+n\log n}$. For unweighted graphs, which we consider in this paper, the heap can be replaced by an array list, which reduces the time complexity to $\cmp{m}$. As an example, the left graph in~\figref{trees} shows a spanning tree computed with Prim's algorithm.

\subsection*{Kruskal's algorithm}

Kruskal's algorithm differs conceptually from Prim's algorithm. Instead of starting with a tree consisting of a seed node and then expanding it, the algorithm starts with a forest of trees, each consisting of a single node. The trees are then incrementally merged into larger trees by adding edges between them until only one remains. At this point the algorithm stops and the remaining tree is a spanning tree of a graph. Since the algorithm turns out inefficient for our purposes in this paper, we do not provide the exact pseudocode here.

Kruskal's algorithm is nondeterministic and the actual spanning tree depends on a random selection of the edges to merge the trees at each step. The time complexity of the algorithm using a disjoint-set data structure is $\cmp{m\log n}$, for either weighted or unweighted graphs. As an example, the middle graph in~\figref{trees} shows a spanning tree computed with Kruskal's algorithm.

\subsection*{Breadth-first search}

The breadth-first search node traversal is very similar to Prim's algorithm. The main difference is in how the edges to non-visited nodes are processed. In contrast to Prim's algorithm, where only one such edge is processed on each step, the breadth-first search processes all edges from a selected node to non-visited nodes in a single step.

The breadth-first search algorithm for computing a spanning tree $T$ of an undirected connected graph $G$ operates as follows (see~\algref{bfs}). In contrast to before, we make all computations in~\algref{bfs} explicit. First, the algorithm selects a random seed node $i\in V_G$ from graph $G$ and adds it to an empty tree $T$ (lines 3, 4). The node $i$ is also added to an empty queue $Q\subseteq V_T$. Then, on each step of the algorithm (lines 5-11), a node $i\in Q$ is removed from the beginning of the queue $Q$ (line 6) and all edges that lead from node $i\in V_T$ already in the tree $T$ to nodes $j\notin V_T$ not yet in the tree $T$ are processed (lines 7-10). All nodes $j$ and edges $\{i,j\}$ are added to the tree $T$ (lines 8, 9), while nodes $j$ are also added to the queue $Q$ for further processing. Finally, when there is no other node $i\in Q$ in the queue $Q$, the algorithm stops (line 5). At this point the tree $T$ is a spanning tree of graph $G$ (line 12).

\begin{algorithm}[h]
	{\small\caption{\label{alg:bfs}Breadth-first search}
	\begin{algorithmic}[1]
		\Require undirected graph $G$
		\Ensure spanning tree $T$
		\State $T\gets$ empty graph
		\State $Q\gets$ empty queue
		\State $i\gets \Call{Random}{i\in V_G}$ \Comment{{\small Random seed node.}}
		\State add node $i$ to $V_T$ and $Q$ \Comment{{\small Add selected seed node.}}
		\While{$\exists i\in Q$} \Comment{{\small There exists non-processed node?}}
				\State $i\gets$ remove node from $Q$ \Comment{{\small Select first non-processed node.}}
		\For{$\{i,j\}\in E_G: j\notin V_T$} \Comment{{\small Edges to non-visited nodes.}}
			\State add node $j$ to $V_T$ and $Q$ \Comment{{\small Add non-visited node.}}
			\State add edge $\{i,j\}$ to $E_T$ \Comment{{\small Add selected edge.}}
		\EndFor
		\EndWhile
		\State \Return $T$
	\end{algorithmic}}
\end{algorithm}


The breadth-first search algorithm is again nondeterministic, while the actual spanning tree depends on a random selection of the seed node (line 3) and on the exact order in which the edges to expand the tree are processed (line 7). The time complexity of the algorithm using a queue of non-processed nodes is $\cmp{m}$, for either weighted or unweighted graphs. As an example, the right graph in~\figref{trees} shows a spanning tree computed with the breadth-first search algorithm.

\subsection*{Other algorithms}

Other algorithms for computing a spanning tree include Sollin's algorithm, which can be seen as a combination of Prim's and Kruskal's approaches, Borůvka's algorithm and the depth-first search node traversal. While the breadth-first search algorithm processes nodes of a graph using a level-order traversal, the depth-first search algorithm uses a preorder traversal. This means that the only change required to the breadth-first search algorithm is to replace the queue of non-processed nodes $Q$ with a stack (line 2 in~\algref{bfs}). 
The time complexity of the depth-first search algorithm is again $\cmp{m}$.

\subsection*{Sackin's index}

In the language of computational theory, a balanced tree is a data structure where the time complexity of standard operations such as adding or deleting an element is $\cmp{\log n}$~\cite{Knu11}. Consider a rooted tree $T$ with root $r\in V_T$ and let $\widetilde{V}_T\subseteq V_T\setminus\{r\}$ be the set of tree leaves (\ie degree-$1$ nodes). Then, balance implies that the distance $d_{ir}$ between all leaf nodes $i\in \widetilde{V}_T$ and the root $r\in V_T$ is at most $\cmp{\log n}$, which further implies that the average distance between all pairs of nodes $\avg{d}$ and also the diameter $d_{\rm max}$ are in $\cmp{\log n}$, since one can always take a path through the root.

Phylogenetics literature defines various indices of tree balance~\cite{LLMN22,FHKKW23} that quantify the branching symmetry and compactness of trees. One of the most widely used is Sackin's index of imbalance~\cite{Sac72}. Sackin's index is defined as the sum of the number of nodes between all leaves $i\in \widetilde{V}_T$ and the root $r\in V_T$, which is included in the count. This can be equivalently written as $\sum_{i\in \widetilde{V}_T}d_{ir}$. Since Sackin's index tends to increase with the number of nodes, we normalize by the minimum possible value $\widetilde{n}=|\widetilde{V}_T|$, which is reached on the star tree, and the maximum possible value $(\widetilde{n}+2)(\widetilde{n}-1)/2$, which is reached on the caterpillar tree. The normalized Sackin's index $S$~\cite{SS90} is then defined as
\begin{eqnarray}
	S=\frac{\sum_{i\in \widetilde{V}_T}d_{ir}-\widetilde{n}}{(\widetilde{n}+2)(\widetilde{n}-1)/2-\widetilde{n}},
\label{eq:sackin}
\nonumber
\end{eqnarray}
where smaller values correspond to more balanced trees.

For spanning trees computed with the breadth-first search and Prim's algorithms, we designate the randomly selected seed node as the root. Spanning trees computed with Kruskal's algorithm, on the other hand, do not have a naturally defined root. We, therefore, randomly select different nodes as the root and average the results.

\section*{Results}
\label{sec:results}

\subsection*{Synthetic graphs}
\label{sec:graphs}

Consider an Erd\H{o}s-R\'{e}nyi random graph~\cite{ER59} with $n$ nodes and the probability of an edge between each pair of nodes $p=\avg{k}/(n-1)$, where $\avg{k}$ is the expected node degree. A spanning tree of any connected graph with $n$ nodes consists of $n$ nodes and $n-1$ edges. Thus, the average node degree is $\avg{k}=2-2/n$. Since it is a tree, the average clustering coefficient is equal to $\avg{C}=0$~\cite{WS98}. Therefore, we focus on other graph properties here. In particular, we study the average distance between the nodes $\avg{d}$ and the diameter $d_{\rm max}$. A theoretical estimate for the diameter $d_{\rm max}$ of a random graph equals $\log n/\log\avg{k}$~\cite{New18c}, which is $d_{\rm max}=2.40$ for $n=250$ and $\avg{k}=10$. Due to the sensitivity of the diameter $d_{\rm max}$ for relatively small $n$ and $\avg{k}$, this turns out to be a better estimate of the average distance between the nodes $\avg{d}$. Indeed, the empirical estimate for the considered random graph is $\avg{d}\approx 2.64$ whereas $d_{\rm max}\approx 4.39$.

\figref{trees} shows particular realizations of spanning trees of a random graph with the above parameters computed using Prim's algorithm, Kruskal's algorithm and the breadth-first search algorithm. The diameter $d_{\rm max}$ of the spanning trees is equal to $14$, $17$ and $6$, respectively. While the diameters of the spanning trees computed with Prim's and Kruskal's algorithms are much higher than in the random graph, the diameter of the spanning tree computed with the breadth-first search algorithm is very close. These observations are closely related to the question of whether the computed spanning trees are balanced.

The average distance $\avg{d}$ and the diameter $d_{\rm max}$ of a balanced tree are in $\cmp{\log n}$ for any practical definition of balance~\cite{Knu11}. However, in the case of a random tree, both values are almost certainly in $\cmp{\sqrt{n}}$~\cite{RS67,MM70}. Since these results only talk about the scaling, they can not be directly employed to measure whether a particular spanning tree is balanced or not. 
Nevertheless, one can study the scaling of the average distance $\avg{d}$ and the diameter $d_{\rm max}$ of spanning trees of graphs with an increasing number of nodes $n$ and empirically estimate whether the values scale as $\cmp{\log n}$ or worse. Note that only in the case of the former the spanning trees can possibly retain short distances between the nodes in random graphs and real small-world networks~\cite{WS98}.

\begin{figure}[t]
\centering\vskip8pt\includegraphics[width=\textwidth]{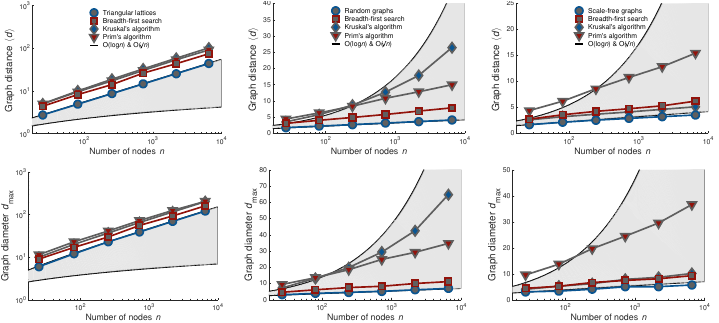}\vskip8pt%
\caption{{\bf The average distance $\avg{d}$ and the diameter $d_{\rm max}$ of triangular lattices (left), random graphs (middle) and scale-free graphs (right), and their spanning trees computed with different algorithms.}
The plots show estimates over $100$ realizations, where the shaded areas span between theoretical estimates for random graphs $\cmp{\log{n}}$ and two-dimensional lattices $\cmp{\sqrt{n}}$, and are consistent between the plots.}
\label{fig:graphs:distance}
\end{figure}

Besides Erd\H{o}s-R\'{e}nyi random graphs~\cite{ER59}, we also analyse triangular lattices and Barab\'{a}si-Albert scale-free graphs~\cite{BA99}. We vary the number of nodes $n$, while we keep the average node degree fixed to $\avg{k}=10$.\footnote{Realizations of synthetic graphs analyzed in the paper are available as Pajek files at \url{https://doi.org/10.5281/zenodo.15034997}.} \figref{graphs:distance} shows the scaling of the average distance $\avg{d}$ and the diameter $d_{\rm max}$ for synthetic graphs and their spanning trees computed with different algorithms.

We first consider triangular lattices, as these results serve as a baseline for further analysis. The average distance $\avg{d}$ and the diameter $d_{\rm max}$ of any two-dimensional lattice scale as $\cmp{\sqrt{n}}$~\cite{New18c}. This can be observed as a straight line with slope $0.5$ on double logarithmic plots in the left column of~\figref{graphs:distance}. Notice that all spanning trees computed with different algorithms show similar scaling $\cmp{\sqrt{n}}$.

\begin{figure}[b]
\centering\vskip8pt\includegraphics[width=0.66\textwidth]{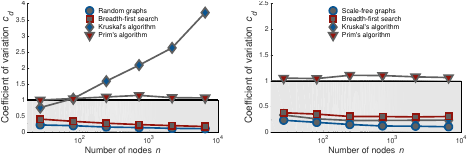}\vskip8pt%
\caption{{\bf The coefficient of variation $c_d$ for random graphs (left) and scale-free graphs (right), and their spanning trees computed with different algorithms.}
The plots show estimates over $100$ realizations, while the error bars are smaller than the symbol sizes.}
\label{fig:graphs:cov}
\end{figure}

Next, we consider Erd\H{o}s-R\'{e}nyi random graphs~\cite{ER59} shown in the middle column of~\figref{graphs:distance}. The average distance $\avg{d}$ and the diameter $d_{\rm max}$ of random graphs, and also small-world networks~\cite{WS98}, scale as $\cmp{\log n}$~\cite{New18c}. This can be observed as a straight line on semi-logarithmic plots in~\figref{graphs:distance}, whereas any upward concave function would imply a faster scaling than $\cmp{\log n}$. Notice that the spanning trees computed with the breadth-first search algorithm best preserve the distances in random graphs, while both the average distance $\avg{d}$ and the diameter $d_{\rm max}$ appear to scale as $\cmp{\log n}$, at least for a moderate number of nodes $n\leq 10^4$. In contrast, the distances between the nodes of the spanning trees computed with Kruskal's algorithm scale faster than $\cmp{\log n}$ (see also~\figref{graphs:cov} and the discussion alongside).

Last, we consider Barab\'{a}si-Albert scale-free graphs~\cite{BA99} shown in the right column of~\figref{graphs:distance}. The average distance $\avg{d}$ and the diameter $d_{\rm max}$ of scale-free graphs scale as $\cmp{\log n/\log\log n}$~\cite{CH03}, while such graphs are usually called ultra small-world~\cite{Bar16}. Note that $\cmp{\log n/\log\log n}$ is indistinguishable from $\cmp{\log n}$ for $n\leq 10^4$, thus this scaling can again be observed as a straight line on semi-logarithmic plots in~\figref{graphs:distance}. The spanning trees computed with both the breadth-first search algorithm and Kruskal's algorithm well retain the distances between the nodes of scale-free graphs and appear to scale as $\cmp{\log n}$. On the other hand, the distances between the nodes of the spanning trees computed with Prim's algorithm can be more than five times larger than the distances in scale-free graphs (\eg $\avg{d}=15.42$ and $d_{\rm max}=36.60$ compared to $3.51$ and $5.64$ for graphs with $n=6\,561$ nodes).

The above observations are confirmed in~\figref{graphs:cov}, where we show the coefficient of variation of the distances between the nodes $c_d$. Note that the distributions of the distances between the nodes in random and scale-free graphs, and real small-world networks, are low-variance with $c_d\ll 1$~\cite{WS98,Bar16}. As one can observe in~\figref{graphs:cov}, all distributions of the distances in the spanning trees computed with the breadth-first search algorithm are low-variance $c_d\ll 1$. On the other hand, the distributions in the spanning trees computed with Kruskal's algorithm are high-variance $c_d\gg 1$ for random graphs with $n>100$, while the results for Prim's algorithm are inconclusive $c_d\approx 1$.

To summarize, if a spanning tree should retain short distances between the nodes of a graph, then the breadth-first search algorithm is the preferred choice, at least for random and scale-free graphs. In the following section, we also consider real networks.

\subsection*{Real networks}
\label{sec:networks}

\tblref{networks} shows statistics of collections of more than a thousand real networks analyzed in the paper. These represent citations between the papers published in the journal Physical Review E~\cite{APS}, paper collaborations between Slovenian researchers extracted from the SICRIS database~\cite{SICRIS}, protein interactions of different species collected from the BioGRID repository~\cite{SBRBBT06,BioGRID}, interactions between the users at the stack exchange web site MathOverflow~\cite{PBL17,SNAP}, Facebook friendships between the students at different US universities~\cite{TMP12,NDR} and links between autonomous systems extracted by the Oregon Route Views project~\cite{LKF05,SNAP}. Some collections represent temporal networks that grow through time (\eg paper citations and author collaborations), while other represent similar networks of different size (\eg protein interactions and online friendships). All networks were reduced to a simple graph of their largest connected component.\footnote{Real networks analyzed in the paper are available as Pajek files at \url{https://doi.org/10.5281/zenodo.15034997}.}

\begin{table}[h]
\begin{adjustwidth}{-2.25in}{0in}
\centering
\caption{
{\bf Statistics of collections of real networks.}}
\begin{tabular}{|l+l|l|l|l|l|l|}
\hline
{\bf Collection} & {\bf Networks $N$} & {\bf Nodes $n$} & {\bf Edges $m$} & {\bf Degree $\avg{k}$} & {\bf Distance $\avg{d}$} & {\bf Clustering $\avg{C}$}\\ \thickhline
Paper citations & $46$ & $[3,37\,511]$ & $[2,135\,260]$ & $[1.3,7.2]$ & $[1.33,21.79]$ & $[0.00,0.27]$ \\ \hline
Author collaborations & $25$ & $[18,1\,735]$ & $[42,6\,710]$ & $[4.1,7.7]$ & $[1.85,8.75]$ & $[0.46,0.75]$ \\ \hline
Protein interactions & $40$ & $[5,19\,961]$ & $[4,238\,886]$ & $[1.6,83.1]$ & $[1.47,6.06]$ & $[0.00,0.52]$ \\ \hline
User interactions & $75$ & $[2,20\,969]$ & $[1,86\,137]$ & $[1.0,10.1]$ & $[1.00,3.80]$ & $[0.00,0.17]$ \\ \hline
Online friendships & $97$ & $[762,41\,536]$ & $[16\,651,1\,590\,651]$ & $[39.1,116.2]$ & $[2.24,3.21]$ & $[0.19,0.41]$ \\ \hline
Autonomous systems & $733$ & $[103,6\,474]$ & $[239,12\,572]$ & $[3.4,4.7]$ & $[2.65,3.98]$ & $[0.16,0.29]$ \\ \hline
\end{tabular}
\begin{flushleft}The statistics include the number of networks in the collection $N$, the number of nodes $n$ and edges $m$, the average node degree $\avg{k}$, the average distance between the nodes $\avg{d}$ and the average node clustering coefficient $\avg{C}$.
\end{flushleft}
\label{tbl:networks}
\end{adjustwidth}
\end{table}

\begin{figure}[p]
\centering\vskip8pt\includegraphics[width=\textwidth]{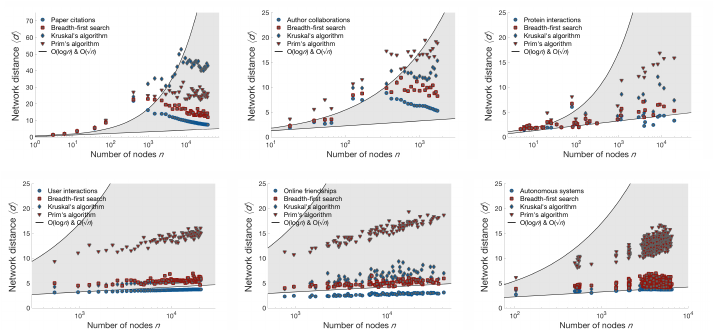}\vskip8pt%
\caption{{\bf The average distance $\avg{d}$ in real networks and their spanning trees computed with different algorithms.}
The shaded areas are the same as in~\figref{graphs:distance}.}
\label{fig:networks:distance}
\end{figure}

\begin{figure}[p]
\centering\vskip8pt\includegraphics[width=\textwidth]{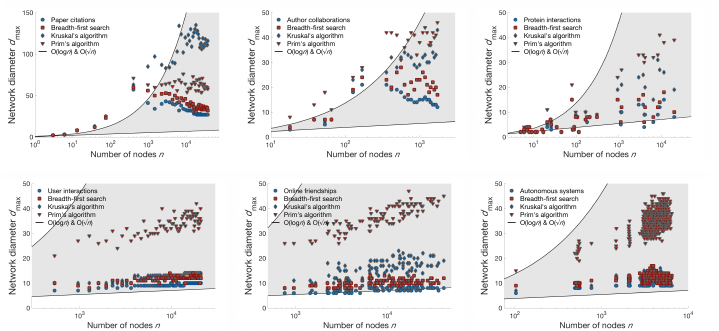}\vskip8pt%
\caption{{\bf The diameter $d_{\rm max}$ of real networks and their spanning trees computed with different algorithms.}
The shaded areas are the same as in~\figref{graphs:distance}.}
\label{fig:networks:diameter}
\end{figure}

\figref{networks:distance} shows the average distance between the nodes $\avg{d}$ in real networks and their spanning trees, where we have used semi-logarithmic axes as in~\figref{graphs:distance}. First, we consider the networks. As expected for small-world networks~\cite{WS98}, the average distance $\avg{d}$ increases with the number of nodes $n$ and appears to scale no faster than $\cmp{\log n}$ in all network collections but two. In the case of temporal networks representing paper citations and author collaborations in the first two plots of~\figref{networks:distance}, the average distance $\avg{d}$ actually starts to decrease when the number of nodes exceeds $n\approx 500$. This is a consequence of network densification known as a shrinking diameter~\cite{LKF05}. \figref{networks:diameter} shows also the diameter $d_{\rm max}$ of real networks, where the interpretation is exactly the same.

Next, we consider the spanning trees of these networks computed with different algorithms. Consistent with the results for synthetic graphs, the spanning trees computed with the breadth-first search algorithm best preserve the average distance between the nodes $\avg{d}$ in all network collections but two. In the case of networks representing user interactions and autonomous systems in the bottom row of~\figref{networks:distance}, Kruskal's algorithm performs similarly well. Furthermore, in non-temporal networks that are not subject to the densification law~\cite{LKF05}, the average distance $\avg{d}$ of the spanning trees computed with the breadth-first search algorithm appears to scale no faster than $\cmp{\log n}$, while in other networks, the scaling of the average distance $\avg{d}$ closely follows the scaling in real networks. Again, \figref{networks:diameter} shows also the diameter $d_{\rm max}$ of spanning trees, where the interpretation is exactly the same.

\begin{figure}[t]
\centering\vskip8pt\includegraphics[width=\textwidth]{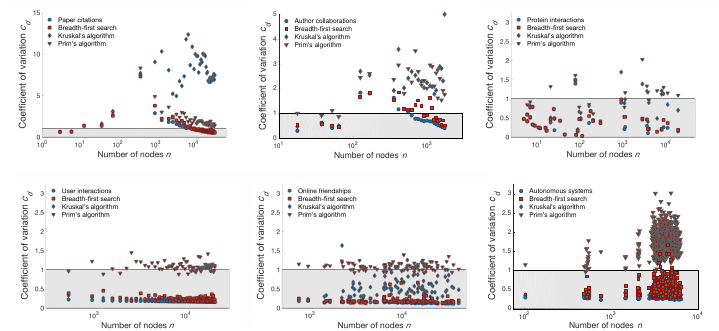}\vskip8pt%
\caption{{\bf The coefficient of variation $c_d$ for real networks and their spanning trees computed with different algorithms.}
Other details are the same as in~\figref{graphs:cov}.}
\label{fig:networks:cov}
\end{figure}

The above observations are confirmed in~\figref{networks:cov}, where we show the coefficient of variation of the distances between the nodes $c_d$. Notice that all distributions of the distances in real networks and spanning trees computed with the breadth-first search algorithm are low-variance with $c_d<1$, as long as the networks are large enough $n\geq 10^4$. In contrast, this holds neither for Kruskal's algorithm nor Prim's algorithm where even $c_d\gg 1$ for some network collections (see first and last plot of~\figref{networks:cov}).

\begin{figure}[p]
\centering\vskip8pt\includegraphics[width=\textwidth]{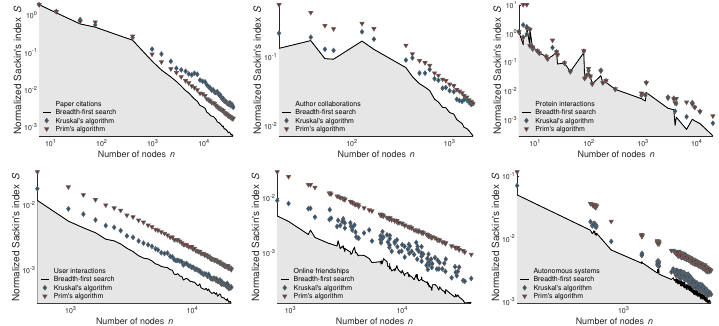}\vskip8pt%
\caption{{\bf Normalized Sackin's index $S$ of spanning trees computed with different algorithms.} The plots show estimates over $25$ realizations.}
\label{fig:networks:sackin}
\end{figure}

A well-balanced spanning tree would already imply short distances between the nodes $\avg{d}\sim \log n$ and also the diameter $d_{\rm max}\sim \log n$. One of the most commonly used measures of tree balance in the phylogenetics literature is Sackin's index of tree imbalance~\cite{Sac72}. \figref{networks:sackin} shows normalized Sackin's index $S$~\cite{SS90} of spanning trees computed with different algorithms. Notice that in all but a few cases, Sackin's index $S$ of spanning trees computed with the breadth-first search algorithm is strictly smaller than that of spanning trees computed with any other algorithm, often by an order of magnitude. Therefore, the breadth-first search algorithm computes the most balanced spanning trees. Another widely used measure of tree compactness or density~\cite{OLW20} from the chemical graph theory literature is the so-called Wiener's index~\cite{Wie47}. Weiner's index is defined as the unnormalized distance between all pairs of nodes ${n \choose 2}\avg{d}$ and thus can be readily observed in~\figref{networks:distance}.

\begin{figure}[p]
\centering\vskip8pt\includegraphics[width=\textwidth]{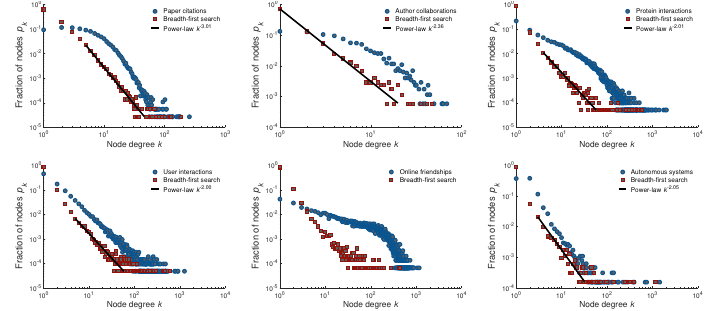}\vskip8pt%
\caption{{\bf Node degree distribution $p_k$ of real networks and their spanning trees computed with the breadth-first search algorithm.}
The power-law distributions are maximum likelihood estimates at $\pval=0.1$~\cite{CSN09}.}
\label{fig:networks:degrees}
\end{figure}

It is further interesting that the node degree distribution $p_k$ of the spanning trees computed with the breadth-first search algorithm often follows a power-law $p_k\sim k^{-\gamma}$~\cite{CSN09}, regardless of whether the network is scale-free or not~\cite{BA99,BC19}. \figref{networks:degrees} shows the node degree distribution $p_k$ of the largest networks in~\tblref{networks} and their spanning trees. Under the goodness-of-fit test at $\pval=0.1$~\cite{CSN09}, a power-law $p_k\sim k^{-\gamma}$ is a plausible fit of the degree distribution $p_k$ for the protein interactions and autonomous systems networks in the right column of~\figref{networks:degrees}. On the other hand, the degree distribution $p_k$ of the spanning trees follows a power-law in all cases but the online friendships network, while the maximum likelihood estimates of the power-law exponents $\gamma$ are shown with solid lines in~\figref{networks:degrees}. Considering the impact of a power-law degree distribution on network structure and dynamics, this property may be useful in practical applications of spanning trees.

Finally, \tblref{complexity} shows the runtime and memory consumption of an efficient {\it Java} implementation of the breadth-first search algorithm and Prim's algorithm applied to the largest networks in~\tblref{networks}. The breadth-first search algorithm is consistently faster than Prim's algorithm and consumes much less memory (up to $20$ times). On the other hand, Kruskal's algorithm is considerably less efficient and is therefore not included in the analysis.

\begin{table}[h]
\begin{adjustwidth}{-2.25in}{0in}
\centering
\caption{
{\bf The runtime and memory consumption of algorithms for computing spanning trees of real networks.}}
\begin{tabular}{|l+l|l|l|l|l|l|}
\hline
 & \multicolumn{3}{|l|}{\bf Wall time $[\mathrm{ms}]$} & \multicolumn{3}{|l|}{\bf Memory $[\mathrm{MB}]$} \\ \hline
{\bf Network} & {\bf Breadth-first srch.} & {\bf Prim's alg.} & {\bf Speed-up} & {\bf Breadth-first srch.} & {\bf Prim's alg.} & {\bf Gain} \\ \thickhline
Paper citations & $13.33\pm 0.39$ & $62.58\pm 0.80$ & $4.7\times$ & $14.88\pm 0.07$ & $53.89\pm 0.07$ & $3.6\times$ \\ \hline
Author collaborations & $0.97\pm 0.08$ & $2.55\pm 0.13$ & $2.6\times$ & $5.03\pm 0.00$ & $10.05\pm 0.00$ & $2.0\times$ \\ \hline
Protein interactions & $8.22\pm 0.41$ & $108.69\pm 1.16$ & $13.2\times$ & $10.23\pm 0.06$ & $195.33\pm 0.20$ & $19.1\times$ \\ \hline
User interactions & $6.03\pm 0.27$ & $78.61\pm 1.03$ & $13.0\times$ & $9.91\pm 0.03$ & $145.95\pm 1.72$ & $14.7\times$ \\ \hline
Online friendships & $31.71\pm 0.67$ & $340.36\pm 0.98$ & $10.7\times$ & $27.90\pm 0.17$ & $386.37\pm 0.54$ & $13.8\times$ \\ \hline
Autonomous systems & $1.37\pm 0.13$ & $31.16\pm 0.77$ & $22.7\times$ & $3.52\pm 0.02$ & $68.52\pm 0.85$ & $19.5\times$ \\ \hline
\end{tabular}
\begin{flushleft}The values are estimates with standard errors over $100$ realizations.
\end{flushleft}
\label{tbl:complexity}
\end{adjustwidth}
\end{table}

To summarize, we again conclude that if a spanning tree should retain the average distance between the nodes $\avg{d}$ and the diameter $d_{\rm max}$ of a real network, then the breadth-first search algorithm should be used. Whether preserving the distances is actually desired or favorable depends on a specific application, which we consider in the following section.

\section*{Applications}
\label{sec:apps}

A spanning tree can be viewed as a technique for discovering a network backbone or skeleton~\cite{CN17,Sub18a}, with applications in network visualization and link prediction. Moreover, a spanning tree is one of the basic approaches for network simplification or sampling~\cite{LF06,BSB17}. Any computation that can be well approximated from a spanning tree of a network, without the need to apply an algorithm to the entire network, can provide computational benefits. In particular, many network applications require superlinear $\cmp{m\log n}$ or even quadratic algorithms $\cmp{mn}$, where $n$ and $m$ are the number of nodes and edges. These include community detection algorithms~\cite{FH16} and techniques for computing node importance or similarity~\cite{New18c}. In contrast, the computation of a spanning tree using the breadth-first search algorithm has a linear complexity $\cmp{m}$ and therefore does not contribute to the overall time complexity.

In this section, we consider two applications of spanning trees, where it is desired to preserve the distances between the nodes of a network.

\subsection*{Node importance}
\label{sec:apps:importance}

\begin{figure}[t]
\centering\vskip8pt\includegraphics[width=\textwidth]{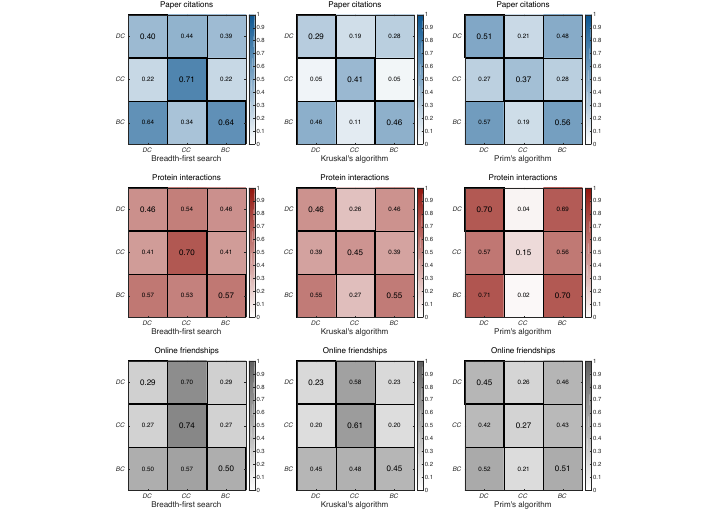}\vskip8pt%
\caption{{\bf Pearson correlation coefficient between node centrality in real networks and their spanning trees computed with the breadth-first search (left), Kruskal's (middle) and Prim's (right) algorithms.}
The measures include node degree centrality $DC$, closeness centrality $CC$ and betweenness centrality $BC$, where the values are estimates over $25$ realizations.}
\label{fig:apps:position}
\end{figure}

Computing the importance of nodes in a network is a classical application of network science with various use cases. There exist many node measures or indices~\cite{SB16}, which are known as measures of node position or centrality in the social network analysis literature~\cite{Fre77,Fre79}. The measure based on the distances between the nodes is called closeness centrality, which estimates the extent to which the node appears to be at the ``center'' of a network. This measure is often used in operations research and logistics. The closeness centrality of a node $i\in V$ in graph $G(V,E)$ can be defined as $\frac{1}{n-1}\sum_{j\neq i}d_{ij}^{-1}$~\cite{Fre79}, where $n=|V|$ is the number of nodes and $d_{ij}$ is the distance between the nodes $i,j\in V$. The time complexity of computing closeness centrality of all nodes in a network is $\cmp{nm}$ and no more efficient algorithm exists~\cite{Knu11}.

\figref{apps:position} shows the Pearson correlation coefficient between node closeness centrality in real networks and their spanning trees computed with the breadth-first search, Kruskal's and Prim's algorithms. The coefficients for the breadth-first search algorithm are shown in the center of the heatmaps in the left column of~\figref{apps:position}. These correlations are all $\geq 0.70$, which indicates a strong linear correlation. On the other hand, the correlation coefficients for the spanning trees computed using Kruskal's algorithm in the middle column and Prim's algorithm in the right column are on average $0.49$ and $0.26$, respectively. Thus, in agreement with the previous results, the breadth-first search algorithm best preserves the distances between the nodes of real networks, not only on average but also at the level of individual nodes.

Merely for comparison, \figref{apps:position} also shows the Pearson correlation coefficient for node degree centrality and betweenness centrality~\cite{Fre77}. The latter measures the extent to which a node appears to serve as a ``bridge'' in a network and is defined as the proportion of all shortest paths between pairs of nodes that pass through a node. The time complexity of computing the betweenness centrality of all nodes in a network is again $\cmp{nm}$~\cite{Bra01}.

\subsection*{Network visualization}
\label{sec:apps:visual}

In addition to the computational benefits, a network simplification technique such as a spanning tree can also be useful for network visualization. Any visualization using a wiring diagram or other approach is limited in the size of a network it can represent and in the structural properties of a network it can reveal~\cite{MM13,GFV13}. Classical algorithms for computing a layout of a network include force-directed algorithms~\cite{Ead84,FR91} and algorithms that embed the nodes in a plane so that their Euclidean distance matches their network distance as closely as possible~\cite{KK89}. It is, therefore, important that a spanning tree preserves the distances between the nodes of a network if it is to be used with such visualization algorithms.

\begin{figure}[t]
\centering\vskip8pt\includegraphics[width=\textwidth]{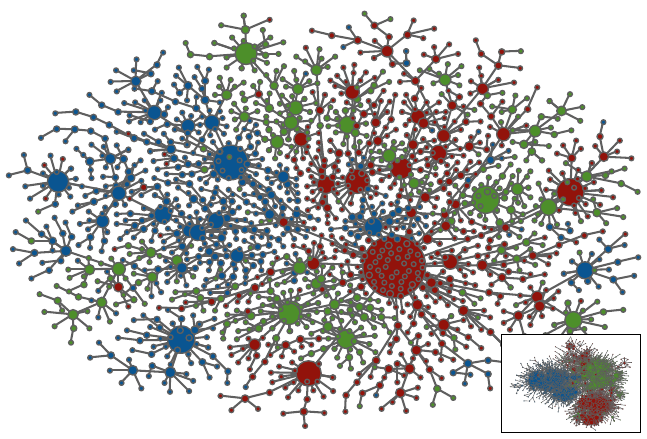}\vskip8pt%
\caption{{\bf Spanning tree of the SICRIS author collaboration network computed with the breadth-first search algorithm.}
The size of the nodes is proportional to their degree, while the colors represent primary author disciplines: natural sciences (red), engineering (green), medical sciences (blue) and others (gray). The inset shows a wiring diagram of the entire network. The layouts were computed with the Large Graph Layout algorithm~\cite{ADWM04}.}
\label{fig:apps:visual}
\end{figure}

\figref{apps:visual} shows a wiring diagram of a spanning tree of the largest connected component of the SICRIS author collaboration network~\cite{SICRIS}. The spanning tree was computed with the breadth-first search algorithm. The wiring diagram illustrates how authors from the same discipline cluster in specific regions and how authors from different disciplines collaborate. In contrast, the visualization of the entire network is much more involved, only revealing three clusters (see inset of~\figref{apps:visual}), while providing very limited insight into the patterns of author collaboration~\cite{SFCK19}. Since there exists no generally accepted quantitative measure for the quality of network visualization, we refrain from further subjective interpretation.

\section*{Conclusion}
\label{sec:conc}

A spanning tree is one of the most straightforward ways to network simplification or sampling and to reveal its backbone or skeleton~\cite{BSB17,CN17}. Well-known algorithms for computing a spanning tree of a weighted network are Prim's and Kruskal's algorithms~\cite{Bol98,New18c}. However, when applied to unweighted networks, which are much more common in practice, these algorithms do not capture the structural properties of real networks, such as short distances between the nodes and a small network diameter. 
On the other hand, an algorithm based on the breadth-first search node traversal well retains the distances between the nodes in synthetic graphs and real networks. The spanning trees are also well-balanced and most compact. As we demonstrate, this can provide computational benefits of up to $20$ times gain and is important for practical applications such as network visualization.
Thus, if a spanning tree of an unweighted network is supposed to retain its structure, then the breadth-first search algorithm should be preferred to other algorithms, and it should be implemented as such in popular network libraries.

We emphasize that the breadth-first search algorithm does not provide the correct result in weighted networks, where Prim's algorithm or Kruskal's algorithm should be used. Furthermore, a spanning tree is not the most suitable technique for simplifying a network in all applications. For instance, a convex skeleton~\cite{Sub18a} is a natural generalization of a spanning tree that also preserves local density, resulting in a high clustering coefficient and an emphasized community structure. 
Moreover, if one is interested in predicting the behavior of dynamical processes, then a high-salience skeleton~\cite{GTB12} might be preferred.

\section*{Acknowledgments}

The authors thank Luka Kronegger for sharing the SICRIS data. This work has been supported by the Slovenian Research Agency ARRS under the program P5-0168.

\nolinenumbers


\end{document}